\documentclass[aps,preprint,amsmath,amssymb,nofootinbib]{revtex4}

\usepackage{hyperref}
\usepackage{graphicx}
\usepackage{slashed}

\begin{document}

\title{Baryogenesis from leptomesons}

\author{Dmitry Zhuridov}%
\email{dmitry.zhuridov@gmail.com}%

\affiliation{Institute of Physics, University of Silesia, Uniwersytecka 4, 40-007, Katowice, Poland}

\date{\today} 

\begin{abstract}
Various new physics models, e.g., theories of compositeness, can accommodate the color singlet excited leptons 
that interact with the leptons, quarks, leptoquarks, etc. 
A particular type of excited lepton, which at low energies interacts with the Standard Model fermions mainly through the four-fermion coupling to lepton and 
quark-antiquark (or lepton-antilepton) pair, we call leptomeson. 
These new particles may contribute to variety of the experimental anomalies such as the discrepancy in the muon $g-2$. 
We propose that the leptomesons can generate also the baryon asymmetry 
that explains the imbalance in ordinary matter and antimatter in the observable universe. 
We consider the two types of scenarios for this baryogenesis via leptogenesis to occur from either 
leptomeson oscillations or decays.  
Both possibilities do not contradict to the small masses of the observable neutrinos and the proton stability. 
Moreover they can be relevant for the near future collider experiments 
and do not suffer from the gravitino problem. 

\end{abstract}

\maketitle

\section{Introduction}

The Standard Model (SM) in particle physics is in good agreement with the majority of the experimental data. 
However it does not explain some fundamental issues, e.g., the large number of the ``elementary" fermions, and their arbitrary 
masses and mixings, fractional electric charge of the quarks, similarity between the leptons and the quarks  
(analogous three flavors, and similar behavior under the $SU(2)_\text{L}\times U(1)_\text{Y}$ symmetry 
with the same weak coupling), etc. 
The models of compositeness~\cite{Pati:1974yy,Terazawa:1976xx,Neeman:1979wp,Shupe:1979fv,Squires:1980cm,Harari:1980ez,
Fritzsch:1981zh,Barbieri:1981cy} try to solve these problems by introducing the substructure of the SM particles. 
Theories with a colored substructure of the leptons besides frequently discussed $SU(3)_\text{c}$ triplet leptoquarks (LQs) 
and octet leptogluons may include also $SU(3)_\text{c}$ singlet excited leptons, which have larger masses, but same 
lepton numbers as the SM leptons. So the leptons can be ``excited'' to these new heavy states by the interactions with 
other SM particles. 
A particular type of the excited leptons present a hypothetical fermion that effectively 
couples to lepton and the pair of the SM fermion and antifermion. 
This coupling conserves the baryon number ($B$) and does not spoil the stability 
of the proton. We refer to the excited lepton of discussed type as leptomeson (LM).\footnote{Notice that in 
Refs.~\cite{Pitkanen:1990,Pitkanen:1992} the same term ``leptomeson'' was used for the bound states of colored 
excitations of $e^+$ and $e^-$.} In particular, LMs may have the same preon content as lepton-meson pairs.

One example of LM generation can be given in the haplon models~\cite{Fritzsch:1981zh,Fritzsch:2014gta}, 
which are based on the symmetry 
$SU(3)_c\times U(1)_{em}\times SU(N)_h$, and contain the two cathegories of colored preons (haplons): the fermions 
$\alpha^{-1/2}$ and $\beta^{+1/2}$, and the scalars $x^{-1/6}$, $y^{+1/2}$, \dots~
In this framework the preon pairs can compose the SM particles as 
$\nu=(\bar\alpha\bar y)_1$, $d=(\bar\beta\bar x)_3$, $W^-=(\bar\alpha\beta)_1$, etc., and the new heavy composites, 
e.g., LQ $(\bar xy)_{\bar3}$ and leptogluon $(\bar\beta\bar y)_8$, where the subindex indicates $SU(3)_c$ 
representation. However there can exist also 
multipreon LM states such as $\bar\beta\bar x\bar y x$, $\bar\alpha\bar y\bar\beta\bar x\beta x$, etc. 
This possibility gets more points from recent discoveries of the multiquark states~\cite{Aaij:2015tga,Chilikin:2013tch} 
due to the similarity between QCD and haplon dynamics. 
Essentially, LMs can be lighter than the LQs and the leptogluons due to the absence of color dressing. 
Notice that possible contribution to the muon $g-2$ from a particular type of LMs, which can couple 
to a lepton and a meson, was discussed in Ref.~\cite{Zhuridov:2015epj}.

One of the most important observations, which can not be explained within the Big Bang cosmology and 
the SM, is the baryon asymmetry ($\eta_B$) of the 
universe that appears to be populated exclusively with baryonic matter rather than antibaryonic matter~\cite{PDG2014}. 
Possible scenarios of dynamical generation of $\eta_B$ during the evolution of the universe from a hot early 
matter-antimatter symmetric stage are known as the baryogenesis (BG) mechanisms. 
Majority of these scenarios discussed in the literature satisfy the three Sakharov conditions~\cite{Sakharov:1967dj}: 
\begin{itemize}
 \item Violation of $B$ symmetry;
 \item Violation of $C$ and $CP$ symmetries (to produce an excess of baryons over antibaryons);
 \item A departure from thermal equilibrium (since the average of $B$ is zero in equilibrium).
\end{itemize}
Some ``exotic" mechanisms of BG that do not satisfy at least one of these conditions 
were discussed in Refs.~\cite{Cohen:1987vi,Dick:1999je,Kobakhidze:2015zka,Hook:2015foa}. 

The SM does not provide a successful BG due to the lack of $CP$ violation and not strongly 
first order electroweak phase transition (PT)~\cite{Trodden:1998ym} to achieve the departure from thermal equilibrium. 
Though in the economical SM extensions $\eta_B$ can be generated through the thermal leptogenesis (LG) 
mechanism~\cite{Fukugita_Yanagida,Davidson:2008bu} 
where the $L$ asymmetry is produced in the out-of-equilibrium decays of heavy Majorana particles, 
and further the SM sphaleron processes~\cite{KRS,Khlebnikov:1988sr} convert this lepton asymmetry into the baryon one. 
These sphaleron transitions are effective until the electroweak symmetry breaking (EWSB). 

However LG in the supersymmetric generalizations of the SM suffers from the 
gravitino problem~\cite{Khlopov1,Balestra,Khlopov2}, which comes from the too high reheating temperature 
related to the strong lower bound on the right-handed neutrino mass 
(Davidson-Ibarra bound)~\cite{Goldberg:1999hp,Barbieri:1999ma,Hamaguchi:2001gw,Davidson:2002qv}. 
To avoid this problem the resonant mechanisms of LG were 
introduced~\cite{Pilaftsis:1997jf,Pilaftsis:2003gt,Pilaftsis:2005rv,Zhuridov:2011ar,Zhuridov:2012hb,Dev:2014laa,Dev:2014wsa}.
 
In this paper we investigate how LMs may provide successful BG.  
The deviation from thermal equilibrium can occur during production (so-called BG 
from oscillations)~\cite{Akhmedov:1998qx,Asaka:2005pn} as well as  
during freeze-out and decay~\cite{Fukugita_Yanagida}.\footnote{It has been shown in Refs.~\cite{Dev:2014laa,Dev:2014wsa} 
that oscillations and decays of heavy sterile neutrinos are indeed two distinct 
sources for baryogenesis via leptogenesis, unlike some previous claims.} 
Depending on the properties of LMs either one of these two scenarios can be realized in the nature. 
The former case can work for both Dirac and Majorana LM masses, and is of particular interest since it 
can be successful with the LM masses of order of EWSB scale that can be tested nowadays. 
The later case requires Majorana masses of LMs similarly to the 
standard LG~\cite{Fukugita_Yanagida,Luty:1992un,Langacker:1986rj,Buchmuller:1996pa,Chen:2007fv} 
from $SU(2)_\text{L}$ singlet neutrino decays. 
However the important difference is that the Davidson-Ibarra bound on the heavy neutrino masses, which 
comes from their see-saw connection to the light neutrino masses through the Yukawa couplings, 
is not applicable to the considered LM masses.
In result, the LM masses are allowed to be 
much smaller than permitted heavy neutrino mass scale of $M_N\gtrsim10^9$~GeV in the standard LG.

\begin{center}
 \begin{figure}[tb]
 \centering
 	\includegraphics[width=1\textwidth]{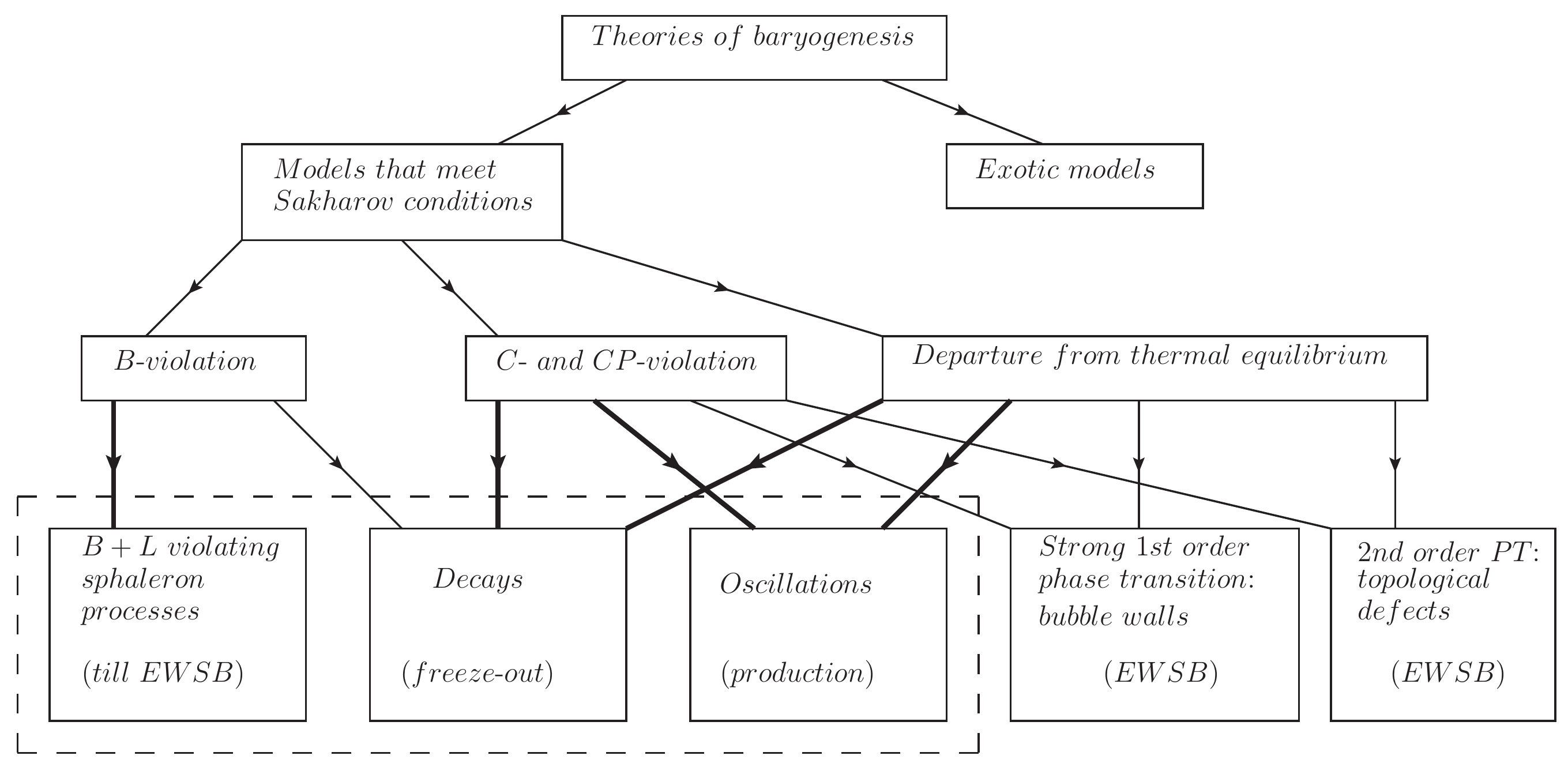} 
	   \caption{Structural scheme for various types of baryogenesis and ways 
	   to meet Sakharov conditions. 
	 }
   \label{Fig:BG}
 \end{figure}
\end{center}

In the flowchart for the BG models shown in Fig.~\ref{Fig:BG} the relevant to present consideration 
ways to satisfy the Sakharov conditions are emphasized by the bold arrows and the related blocks 
are encircled by the dashed line. 
Notice that the models that satisfy these conditions in a non-typical way such as in Ref.~\cite{Fujita:2016igl} 
are not specified in this flowchart. 

In the sections~\ref{section:LG1} and \ref{section:LG2} we present possible BG mechanisms from LM oscillations and decays, 
respectively. We discuss the issue of neutrino masses and conclude in the section~\ref{sec:conclusion}.

\section{Baryogenesis from leptomeson oscillations}\label{section:LG1}

Consider neutral long-living LMs that interact with the SM leptons and quarks at the energies below the new physics scale 
$\Lambda$ (e.g., the compositeness scale) dominantly through 
the effective four-fermion terms, unlike the ordinary sterile neutrinos with their essential Yukawa couplings to the leptons 
and the Higgs doublet. 
For the vector case with lepton flavor and $B$ conservation (and lepton number $L$ conservation for the Dirac LMs) 
these four-fermion interactions at the first order in LM fields can be written as
\begin{eqnarray}\label{eq:4vertex}
    \mathcal{L}_\text{int} = \sum_{\psi_\ell,f,f^\prime} 
    \sum_{\alpha,\beta=L,R}   \left[ 
    \frac{\epsilon^{\alpha\beta}_{ff^\prime\psi_\ell}}{\Lambda^2} 
    (\bar f_\alpha\gamma^\mu f_\alpha^\prime)(\bar\psi_{\ell\beta}\gamma_\mu \ell_{M\beta}^0)   
  + \frac{\tilde\epsilon^{\alpha\beta}_{ff^\prime\psi_\ell}}{\Lambda^2} 
    (\bar\psi_{\ell\alpha}\gamma^\mu f_\alpha^\prime)(\bar f_{\beta}\gamma_\mu \ell_{M\beta}^0)
    \right]
    + \text{H.c.},
\end{eqnarray}
where $\epsilon$ and $\tilde\epsilon$ are the effective couplings 
(real couplings can work for the BG in this section),  
$\psi_\ell=\ell,\nu_\ell$ ($\ell=e,\mu,\tau$) is the SM lepton,
$f$ and $f^\prime$ denote either two quarks or two leptons (we take them from the same 
particle generation) such that the sum of the electric charges of $f_\alpha$, 
$f^{\prime\dag}_\alpha$ and $\psi_{\ell\beta}$ is zero, 
and $\ell_M^0$ is the neutral LM flavor state  that is related to the mass eigenstates $L_{Mi}^0$ 
by the mixing matrix $U$ as
\begin{eqnarray}\label{eq:superposition}
 \ell_{M\alpha}^0 = \sum_{i=1}^n U_{\ell i}^\alpha L_{Mi}^0.
\end{eqnarray}

LMs can be produced thermally from the primordial plasma.  
Once created $\ell_M^0$ oscillate and interact with ordinary matter. 
These processes do not violate the total lepton number $L^\text{tot}$, defined as usual lepton number 
plus that of LMs. However the oscillations of LMs violate $CP$ and therefore their individual lepton numbers 
are not conserved.\footnote{For Majorana LMs the $CP$-violating scatterings can significantly effect the picture 
similarly to the case of sterile neutrinos~\cite{Canetti:2014dka}.} 
Hence the initial state with all zero lepton numbers evolves into a state with $L^\text{tot}=0$ but nonzero individual 
lepton numbers of LMs.

At the temperatures below $\Lambda$ scale 
LMs communicate their lepton asymmetry to the neutrinos and the charged leptons 
through the effective four-fermion interactions in Eq.~\eqref{eq:4vertex}. 
Suppose that the neutral LMs of at least one type remain in 
thermal equilibrium until the time of EWSB $t_\text{EW}$ at which sphalerons become ineffective, and those of at least 
one other type come out-of-equilibrium by $t_\text{EW}$. Hence the lepton number of the former (later) affects 
(has no effect on) the baryogenesis. In result, the final baryon asymmetry after $t_\text{EW}$ 
is nonzero.
At the time $t\gg t_\text{EW}$ all LMs decay into the leptons and the quarks (hadrons).  
For this reason they do not contribute to the dark matter in the universe, and do not destroy 
the Big Bang nucleosynthesis.

The system of $n$ types of singlet LMs with a given momentum $k(t)\propto T(t)$ that interact 
with the primordial plasma can be described by the $n\times n$ density matrix $\rho(t)$. 
In a simplified picture this matrix satisfies the kinetic equation~\cite{Sigl:1992fn,Akhmedov:1998qx}
\begin{eqnarray}\label{eq:evolution}
  i \frac{d\rho}{dt} = [\hat H,\rho] - \frac{i}{2}\{\Gamma,\rho\} + \frac{i}{2}\{\Gamma^p,1-\rho\},
\end{eqnarray}
where $\Gamma$ ($\Gamma^p$) is the destruction (production) rate, and the effective Hamiltonian can be written as 
\begin{eqnarray}
  \hat H = V(t) + U \frac{\hat M^2}{2k(t)} U^\dag,
\end{eqnarray}
where $\hat M^2=\text{diag}(M_1^2,\dots,M_n^2)$ is the matrix of the squared LM mass eigenstates, and $V$ is a real potential. 
(In the approximation of Boltzmann statistics the last term in Eq.~\eqref{eq:evolution} is $i\,\Gamma^p$.) 
In general, evolution of LMs can be considered together with the evolution of the SM leptons 
using the methods of Refs.~\cite{Asaka:2005pn,Asaka:2006rw}. However 
such precise numerical analysis is beyond the scope of present consideration. 
In the following we concentrate on the essentially different temperature dependence of the interaction rate for LMs 
and the sterile neutrinos, which makes the LM scenario more attractive for the experimentalists.

The cross sections for $2\leftrightarrow2$ reactions that contribute to the LM destruction rate can be written as
\begin{eqnarray}\label{eq:cross_section}
  \sigma \equiv \sigma(a+b \leftrightarrow c+d) = 
        \frac{C}{4\pi}\epsilon^2  \frac{s}{\Lambda^4},
\end{eqnarray}
where $a, b, c$ and $d$ denote the four interacting particles 
($f$, $f^\prime$, $\psi_\ell$ and $\ell_M^0$), $C=\mathcal{O}(1)$ is the constant 
that includes the color factor in the case of the interaction with quarks, $s$ is the total energy of the process, 
and $\epsilon$ is the relevant coupling from Eq.~\eqref{eq:4vertex}. 
In the considered LM scenario the cross section in Eq.~\eqref{eq:cross_section}
is proportional to $s$ in contrast to the inverse 
proportionality in the case of BG from neutrino oscillations. 
The respective $2\leftrightarrow2$ scattering rate density at the hight temperatures $M_i \ll T \ll \Lambda$ 
can be calculated as~\cite{Davidson:2008bu}
\begin{eqnarray}
  \gamma_s &=& \frac{g_ag_bT}{32\pi^4} \int_0^\infty ds \, s^{3/2} K_1\left(\frac{\sqrt{s}}{T}\right) \sigma(s) \nonumber\\
	 &=& \frac{6C}{\pi^5}g_ag_b \, \epsilon^2  \frac{T^8}{\Lambda^4},
\end{eqnarray}
where $g_a$ is the number of internal degrees of freedom of the particle $a$, and $K_1$ is the Bessel function. 
Then the interaction rate that equilibrates LMs (average destruction rate) can be estimated as
\begin{eqnarray}\label{eq:rate}
  \Gamma \sim   \epsilon^2  \frac{T^5}{\Lambda^4}.
\end{eqnarray}

The conditions that LMs of type $L_i^0$ remain in thermal equilibrium till the time of the EWSB $t_\text{EW}$, 
while LMs of type $L_j^0$ do not, are
\begin{eqnarray}
  \Gamma_i(T_\text{EW}) &>&  H(T_\text{EW}), \label{eq:Gamma_i} \\
  \Gamma_j(T_\text{EW}) &<&  H(T_\text{EW}), \label{eq:Gamma_j}
\end{eqnarray}
where the Hubble expansion rate $H$ can be written as
\begin{eqnarray}
 H(T) \approx 1.66 g_*^{1/2} \frac{T^2}{M_\text{Planck}},
\end{eqnarray}
where $M_\text{Planck}=1.221\times10^{19}$ GeV is the Planck mass, and $g_*\sim10^2$ is the number of relativistic degrees of freedom in 
the primordial plasma. 

Remarkably, the rates in Eqs.~\eqref{eq:Gamma_i} and \eqref{eq:Gamma_j} are suppressed by the forth power of 
$T_\text{EW}/\Lambda$ ratio with respect to 
the case of the BG via the sterile neutrino oscillations. For this reason, the couplings $\epsilon$ can be 
significantly larger than the Yukawa couplings of that sterile neutrinos. 
In particular, for $\Lambda\gtrsim10$~TeV we have $\epsilon\gtrsim10^{-4}$. Hence the considered scenario of the 
BG via neutral LMs can be relevant for the LHC and next collider experiments without unnatural hierarchy of couplings. 

In the approximation of Eq.~\eqref{eq:evolution} the asymmetry transferred to usual leptons by $t_\text{EW}$ 
can be written as~\cite{Akhmedov:1998qx}
\begin{eqnarray}
  \frac{n_L-n_{\bar L}}{n_\gamma} = \frac{1}{2} \sum_j|S^M_j(t_\text{EW},0)|^2_{CP-\text{odd}},
\end{eqnarray}
where the factor $1/2$ accounts for the photon helicities, and $S^M=U^\dag SU$ is the evolution matrix 
in the mass eigenstate basis ($S(t,t_0)$ is the non-unitary evolution matrix corresponding 
to the operator $\hat H - (i/2)\Gamma$).

In the case of three LM mass states the respective $CP$-violating effects should be proportional to the 
Jarlskog determinant~\cite{Jarlskog:1985ht} related to their mixing matrix $U$. 
However extra LM mass states can enrich the picture of $CP$ violation.
Also additional $CP$-violating phases may come into play from the active neutrino sector 
(compare to Ref.~\cite{Asaka:2005pn}).

\begin{center}
 \begin{figure}[tb]
 \centering
 	\includegraphics[width=0.26\textwidth]{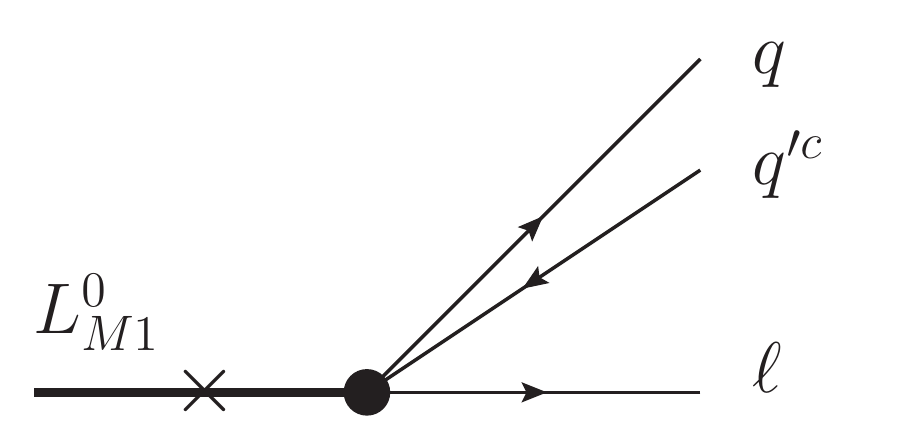} \
 	\includegraphics[width=0.32\textwidth]{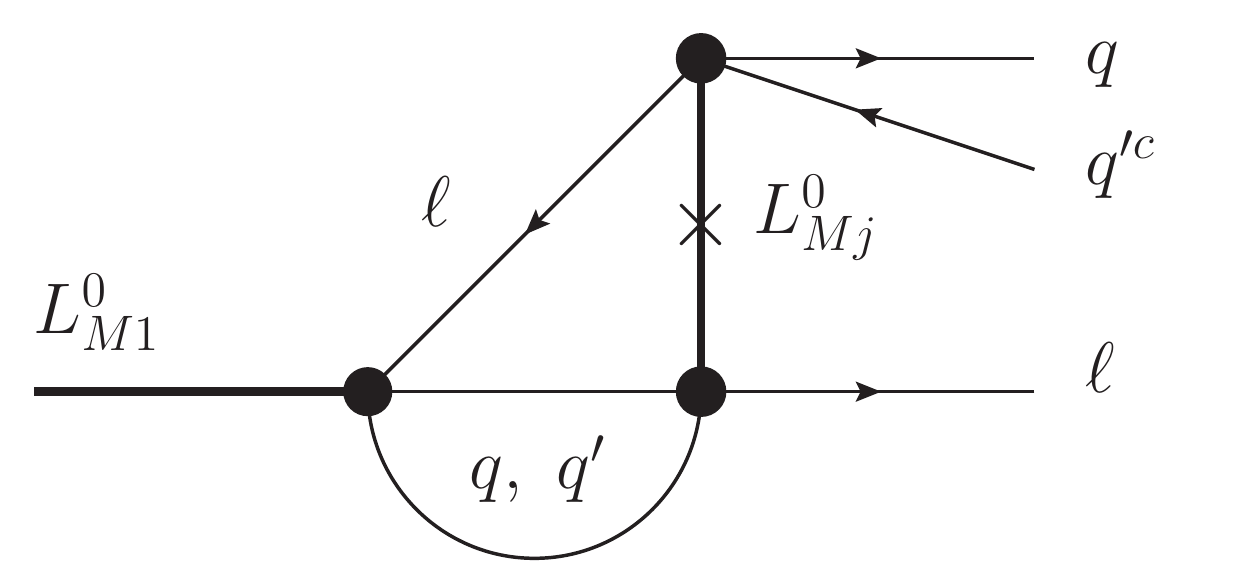} \,  
 	\includegraphics[width=0.37\textwidth]{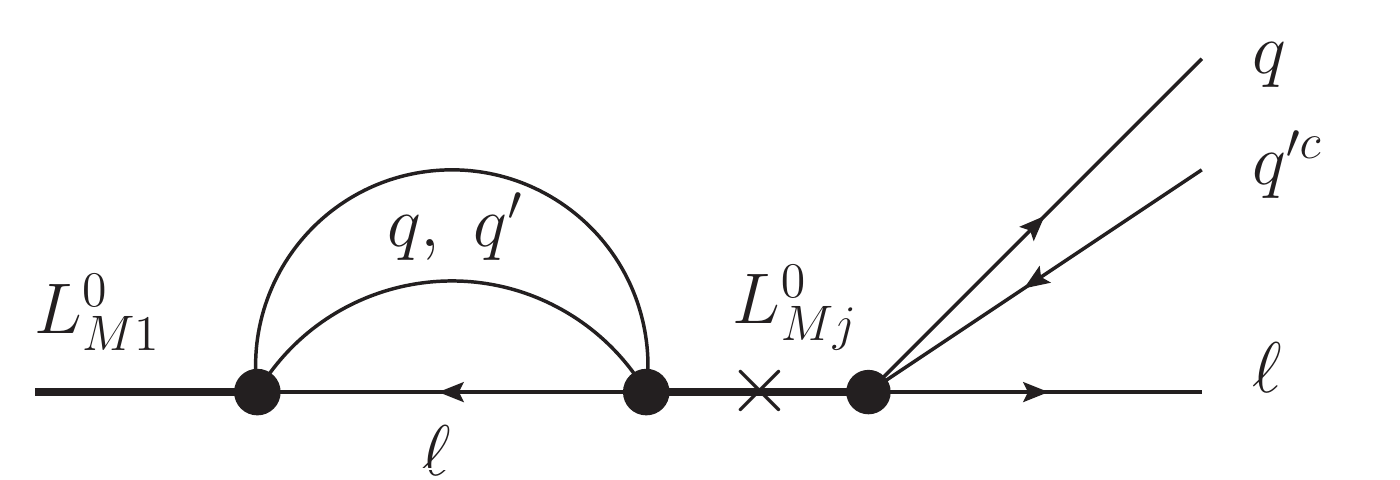}
	   \caption{Feynman diagrams for the discussed contributions to the CP asymmetry, where  
	   the X represents a Majorana mass insertion, the line direction shows either $L$ or 
	   $B$ flow, and the black bulbs represent subprocesses (a particular case of the leptoquark 
	   $S_{0R}$ exchange is shown in Fig.~\ref{Fig:LG_LQ_diagrams} (left)). 
	 }
   \label{Fig:LG_diagrams}
 \end{figure}
\end{center}
\section{Baryogenesis from leptomeson decay}\label{section:LG2}

Suppose that the neutral LMs are Majorana particles ($\ell_{MR}^0=\ell_{MR}^{0c}$). 
Then an analog of usual LG can take place  
due to their out-of-equilibrium, $CP$- and $L$-violating decays 
in the early universe. The relevant terms among the $B$- and lepton flavor-conserving LM interactions 
can be written as
\begin{eqnarray}\label{eq:4vertex_decay}
    \frac{\epsilon_{ff^\prime\psi_\ell}^{\alpha R}}{\Lambda^2} (\bar f_\alpha\gamma^\mu f_\alpha^\prime) 
    (\bar\psi_{\ell R}\gamma_\mu \ell_{MR}^0) + 
    \frac{\epsilon_{ff^\prime\psi_\ell}^{S}}{\Lambda^2} (\bar f_R f_L^\prime) 
    (\bar\psi_{\ell L} \ell_{MR}^0) +
    \frac{\epsilon_{ff^\prime\psi_\ell}^{T}}{\Lambda^2} (\bar f \sigma^{\mu\nu} f^\prime) 
    (\bar\psi_{\ell L} \sigma_{\mu\nu} \ell_{MR}^0) 
    +  \text{H.c.}, 
\end{eqnarray}
where the sum of the hypercharges of $f$, $f^{\prime\dag}$ and $\psi_\ell$ is zero. 
To be more specific in the following we consider the term 
\begin{eqnarray}
    \frac{\lambda_{\ell i}}{\Lambda^2} (\bar q_\alpha\gamma^\mu q_\alpha^\prime) 
    (\bar\ell_R\gamma_\mu L_{Mi}^0),
\end{eqnarray}
where $\lambda_{\ell i}=\epsilon_{qq^\prime\ell}^{\alpha R} U_{\ell i}^R$ is the complex parameter, 
and we used Eq.~\eqref{eq:superposition}. 

Consider the interference of the tree and two-loop diagrams\footnote{Same two-loop self-energy graph was 
discussed in the resonant BG mechanisms of Refs.~\cite{Dev:2015uca,Davoudiasl:2015jja}, where the baryon asymmetry is 
directly produced in the three-body decays of sterile neutrinos $N$. 
Although these mechanisms involve $B$-violating interactions of $QQQN$ type, they do not lead to fast proton decay 
due to the large values of $N$ mass and the $B$-violating interaction scale of $\mathcal{O}(1)$~TeV. These mechanisms 
can be probed in the near future by the neutron-antineutron oscillations and other $B$-violating processes.} 
shown in Fig.~\ref{Fig:LG_diagrams}, 
where $L$ is violated by two units due to the Majorana mass insertion. 
The $CP$ asymmetry that is produced in $L_{M1}^0$ decays can be defined as
\begin{eqnarray}
 \varepsilon_1 = \frac{1}{\Gamma_1}\, \sum_\ell [\Gamma(L_{M1}^{0} \to \ell_R q_\alpha q_\alpha^{\prime c}) 
 - \Gamma(L_{M1}^{0} \to \ell_R^c q_\alpha^c q_\alpha^\prime)],
\end{eqnarray}
where the three-particle decay width is~\cite{Cakir:2002eu}
\begin{eqnarray}
 \Gamma_1 = 
 \sum_\ell [\Gamma(L_{M1}^{0} \to \ell_R q_\alpha q_\alpha^{\prime c}) 
 + \Gamma(L_{M1}^{0} \to \ell_R^c q_\alpha^c q_\alpha^\prime)] \simeq  \frac{1}{128\pi^3}\, 
 (\lambda^\dag \lambda)_{11} \frac{M_1^5}{\Lambda^4}
\end{eqnarray}
with the mass $M_1$ of $L_{M1}^0$. This $CP$ asymmetry to be nonzero requires 
$\text{Im}[ (\lambda^\dag \lambda)_{1j}^2 ] \neq 0$. Hence at least two LM mass states are needed. 
In the case of quasi-degenerate LM masses of $M_2-M_1\sim \Gamma_1/2 \ll M_1$ the self-energy graph gives the dominant 
contribution to the $CP$ asymmetry that can be expressed in the same form~\cite{Dev:2015uca,Davoudiasl:2015jja} 
as in the usual resonant LG~\cite{Pilaftsis:1997jf,Pilaftsis:2003gt,Pilaftsis:2005rv}.
%
In the strong washout regime~\cite{Buchmuller:2004nz} 
the final $B-L$ asymmetry generated at $T\sim M_1$ is insensitive to any initial asymmetry at $T\gg M_1$. 
The respective condition for the decay parameter $K\equiv\Gamma_1 / H(T=M_1) > 3$ translates into the limit of 
\begin{eqnarray}\label{eq:constraint_lambda}
 (\lambda^\dag \lambda)_{11} >  4 \times 10^{-7} \times \left( \frac{\Lambda}{10~\text{TeV}} \right)^4 \times
 \left( \frac{1~\text{TeV}}{M_1} \right)^3.
\end{eqnarray}

The discussed effective LM-quark-antiquark-lepton vertices can be economically realized, e.g.,  
through exchange of the scalar $SU(2)_L$ singlet LQ $S_{0R}$ with the weak hypercharge 
$Y=1/3$~\cite{Buchmuller:1986zs,CiezaMontalvo:1998sk}. 
The relevant interaction terms in the Lagrangian can be written as
\begin{eqnarray}
    -\mathcal{L}_\text{int} &=& (g_{ij}\, \bar d_{R}^c L_{Mi}^0 
	+  f_j\, \bar u^c_R \ell_R ) S_{0R}^j 
	+ \text{H.c.}
\end{eqnarray}
Then the above expressions are valid with the replacements 
$\lambda \to gf^*$ and $\Lambda \to M_{S_{0R}}$. In particular, typical values of the new couplings 
in Eq.~\eqref{eq:constraint_lambda}, e.g., 
$|g|\sim |f|\sim 0.01-0.1$, can be interesting for the collider searches.

\begin{center}
 \begin{figure}[tb]
 \centering
 	\includegraphics[width=0.34\textwidth]{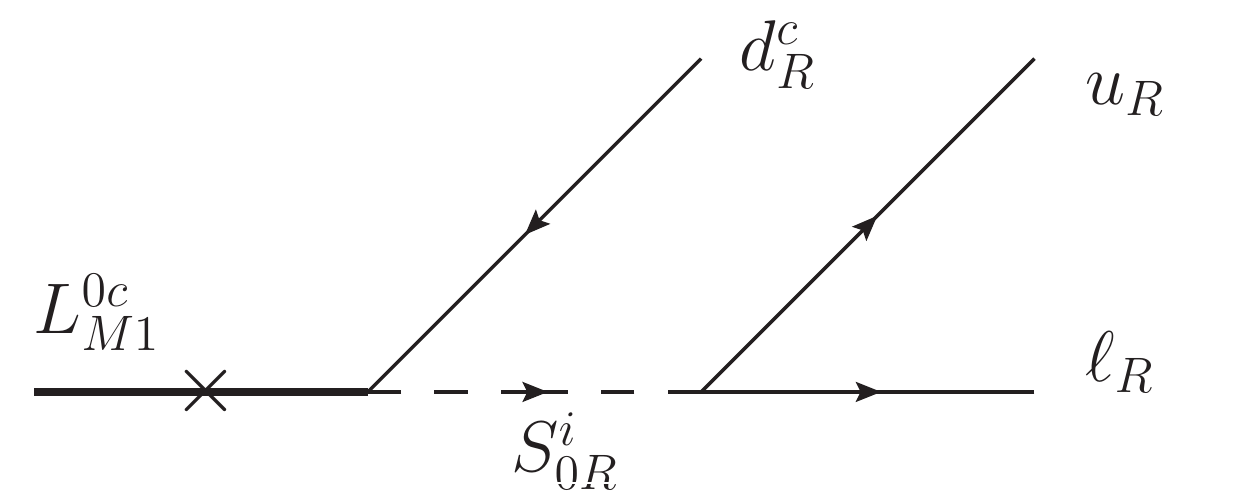} \quad
 	\includegraphics[width=0.5\textwidth]{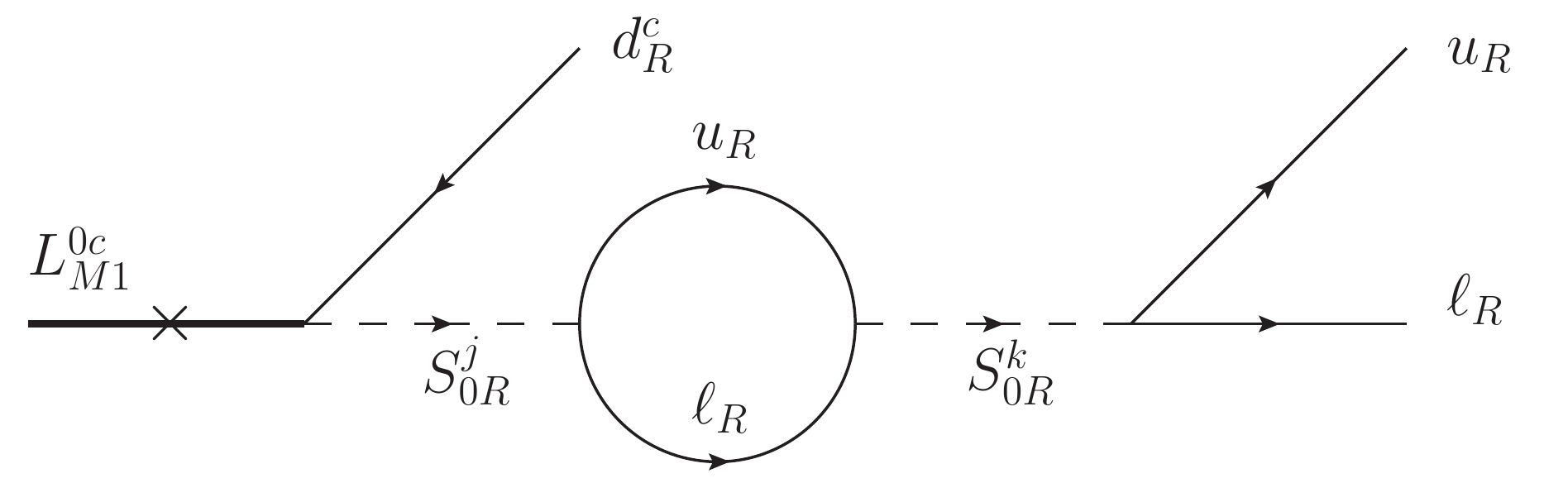}
	   \caption{Discussed Feynman diagrams in the model with scalar leptoquarks $S_{0R}^i$.
	 }
   \label{Fig:LG_LQ_diagrams}
 \end{figure}
\end{center}

Notice that the new contributions to the $CP$ asymmetry coming from the interferences among the tree 
and one-loop diagrams shown in Fig.~\ref{Fig:LG_LQ_diagrams} 
cancel each other unlike the more sophisticated case of Ref.~\cite{Hambye:2001eu} with the three types of 
interactions involved in the LG based on the three body decays. 
However the compositeness models with LQs, which have at least three types of interactions, can realize 
the LG of this kind from LM decays.

The final baryon asymmetry can be written as~\cite{Kolb:1990vq} 
\begin{eqnarray}\label{eq:eta_B}
  \eta_B &\equiv& \frac{n_B-n_{\bar B}}{n_\gamma} = 7.04 \times\frac{n_B-n_{\bar B}}{s} = 
  7.04\times\left(-\frac{28}{79}\right)\times \frac{n_L-n_{\bar L}}{s}  \nonumber\\
         &=& 7.04\times\left(-\frac{28}{79}\right)\times \frac{\varepsilon_1 \kappa}{g_*},
\end{eqnarray}
where $n_B$, $n_L$ and $n_\gamma$ is the baryon, lepton and photon number density, respectively; 
$s$ is the entropy density, $\kappa\leq1$ is the washout coefficient, and $-28/79$ is the sphaleron 
lepton-to-baryon factor. To exactly determine $\kappa$ one should solve 
the set of Boltzmann equations, 
which in case of the resonant regime can be written as
\begin{eqnarray}
    \frac{dN_i}{dz} &=& - (D_i + S_i) (N_i - N_i^\text{eq}), \\
    \frac{dN_{B-L}}{dz} &=& -  \sum_i \varepsilon_i D_i (N_i - N_i^\text{eq}) -  N_{B-L} \sum_i W_i,
\end{eqnarray}
where $z = M_1/T$ is a dimensionless variable, 
$N_X$ [$N^\text{eq}_X$] (with $X=i$ and $B-L$) is the [equilibrium] 
number density of $L_{Mi}^0$ and $B-L$, respectively, and the various reaction rates are denoted by the following factors: 
$D_i$ for $L_{Mi}^0\to\ell qq^{\prime c}$ decays, $S_i$ for the scatterings of $L_{Mi}^0\ell^c \to qq^{\prime c}$, 
$L_{Mi}^0q^{\prime}\to\ell q$, etc., and $W_i$ for the washout processes that include the scatterings and 
the inverse decays of $\ell qq^{\prime c}\to L_{Mi}^0$. It was shown in Ref.~\cite{Dev:2015uca} for the 
similar processes, which are generated by the operator $QQQN$ with a sterile neutrino $N$ 
(instead of $L_M^0\bar L Q\bar Q$ operator in our model), 
that for the interesting 
parameter range of $M\sim1$~TeV and $\Lambda\sim10$~TeV the decay rate 
dominates over the scattering rates at $T\sim T_\text{EW}$ as required for successful BG 
in the strong washout regime. 
%
Using the resonant $CP$ asymmetry of 
\begin{eqnarray}
  \varepsilon_i \sim  \frac{\text{Im}\{[(\lambda^\dag\lambda)_{ij}]^2\}}{(\lambda^\dag\lambda)_{ii}
  (\lambda^\dag\lambda)_{jj}}   \frac{\Gamma_j}{M_j}   \frac{M_iM_j}{M_i^2-M_j^2}  
  \sim \mu^{-1} \frac{\Gamma_1}{M_1}
\end{eqnarray} 
the observed baryon asymmetry 
$\eta_B=(6.21\pm0.16)\times10^{-10}$~\cite{Davidson:2008bu} can be produced 
for the decay parameter of $K\sim100$ and the degeneracy factor of
\begin{eqnarray}
  \mu \equiv \frac{M_2-M_1}{M_1} \lesssim 10^{-6} \left(\frac{M_1}{1~\text{TeV}}\right).
\end{eqnarray}
 
Notice that the dependence of BG on the nonthermal production mechanism for the 
decaying neutral particles responsible for the BG was discussed in Ref.~\cite{Davoudiasl:2015jja} 
for the model based on $QQQN$ operator.

\begin{center}
 \begin{figure}[tb]
 \centering
 	\includegraphics[width=0.3\textwidth]{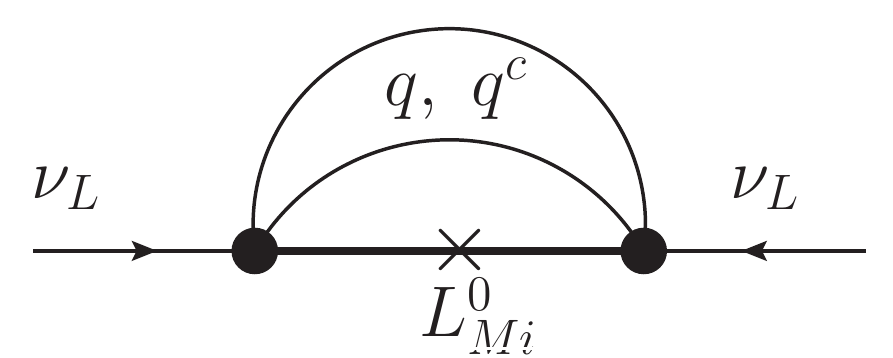} \qquad 
 	\includegraphics[width=0.36\textwidth]{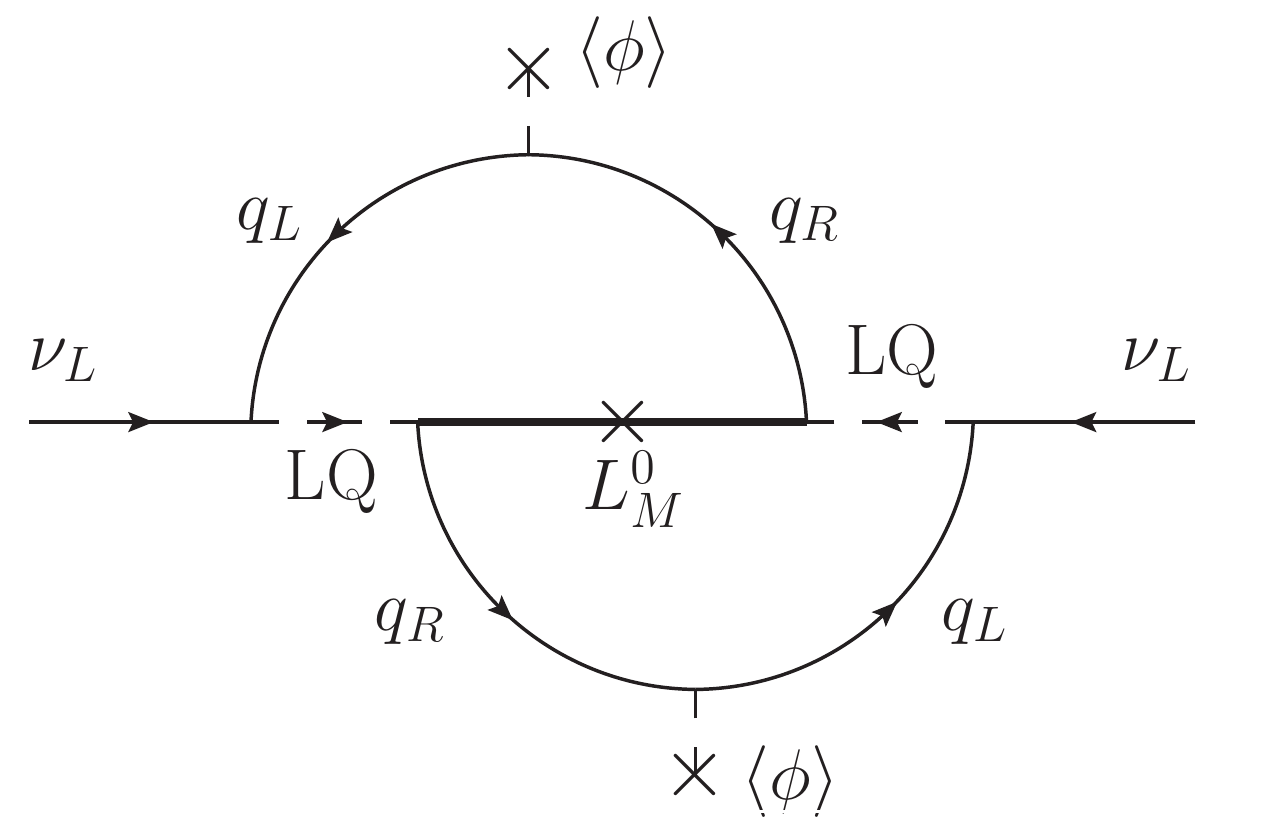}
	   \caption{Discussed diagrams for LM contribution to the neutrino masses. 
	 }
   \label{Fig:nu_mass}
 \end{figure}
\end{center}
\section{Discussion and Conclusion}\label{sec:conclusion}

Substantial feature of any successful BG scenario is consistency with the present bounds on the active neutrino masses. 
In the case of Majorana LMs among the discussed four-fermion interactions 
the terms 
\begin{eqnarray}\label{eq:nu-mass}
      \frac{\epsilon_{ff\nu_\ell}^{S}}{\Lambda^2} (\bar f_R f_L) 
    (\bar\nu_{\ell L} \ell_{MR}^0) 
    +  \frac{\epsilon_{ff\nu_\ell}^{T}}{\Lambda^2} (\bar f \sigma^{\mu\nu} f) 
    (\bar\nu_{\ell L} \sigma_{\mu\nu} \ell_{MR}^0)
    +   
    \text{H.c.}
\end{eqnarray}
can generate a two-loop contribution to the neutrino masses. For $f=q$ this contribution 
can be illustrated by the generic diagram in Fig.~\ref{Fig:nu_mass}~(left), 
where the black bulbs represent some subprocesses. 
Its particular realization in a model with LQs is shown in Fig.~\ref{Fig:nu_mass}~(right). 
The resulting neutrino mass can be estimated as 
\begin{eqnarray}
  m_{\nu_\ell} \sim  \sum_i \frac{(\epsilon\, U_{\ell i}^R)^2}{(16\pi^2)^2}\ \frac{M_i^3 m_f^2}{\Lambda^4},
\end{eqnarray}
where $\epsilon$ is a relevant coupling from Eq.~\eqref{eq:nu-mass}. 
Then present experimental upper bound on the 
neutrino mass of $m(\nu_e)\lesssim2$~eV~\cite{Aseev:2011dq} 
can be easily satisfied for the discussed values of $\epsilon$, $M_i$ and $\Lambda$.

To conclude, we have introduced the two possible generic scenarios of low temperature BG in the new 
class of models with LM states. 
The BG from LM decay can be realized if all LMs decay before the EWSB. In case of relatively light 
and long-lived LMs, which do not all decay before the EWSB the BG from LM oscillations may take place. 
One of the attractive features of this scenario is that the 
out-of-equilibrium condition is more relaxed with respect to the BG from the sterile neutrino oscillations. 
Namely, the constraint on the effective LM coupling $\epsilon$ 
is essentially weakened by the factor of $(\Lambda/T_\text{EW})^2$ with respect to the strong constraints 
on the sterile neutrino Yukawa's. For the contact interactions scale of $\Lambda\sim10$~TeV this factor 
is of $\mathcal{O}(10^3)$ that offers great prospects for the experimental searches of relevant LMs.  
Hence accurate examination of the allowed parameter spaces for the successful BG in the specific models with LMs is 
desirable in the next step.

\section*{Acknowledgements} 

The author thanks Marek Zra{\l}ek, Henryk Czy\.z, Karol Ko\l{}odziej, Tomasz Jeli\'nski and Yue Zhang 
for useful discussions and comments, and the PRD referee for pointing out Refs.~\cite{Dev:2015uca,Davoudiasl:2015jja}. 
This work was supported in part by the Polish National Science Centre, grant number DEC-2012/07/B/ST2/03867. 
The author used JaxoDraw~\cite{Binosi:2003yf} to draw the Feynman diagrams.

\end{document}